\documentclass{elsart3}
\usepackage{graphicx}
\usepackage{amsmath}
\usepackage{amssymb}

\begin{document}

\begin{frontmatter}

\title{Lattice and superexchange effects in doped CMR manganites}

\author[dd]{Alexander Wei{\ss}e},
\author[hgw]{Holger Fehske}$^{,*}$
\corauth{Corresponding Author.\\
  phone: +49-3834-86-4760; fax +49-3834-86-4701.\\
  {\it e-mail:} fehske@physik.uni-greifswald.de}
\address[dd]{Max-Planck-Institut f\"ur Physik komplexer Systeme, N\"othnitzer Stra{\ss}e 38, 01187 Dresden, Germany}
\address[hgw]{Institut f\"ur Physik, Ernst-Moritz-Arndt Universit\"at Greifswald, 17487 Greifswald, Germany}

\begin{abstract}
  We report on the influence of the lattice degrees of freedom on
  charge, orbital and spin correlations in colossal magnetoresistance
  (CMR) manganites. For the weakly doped compounds we demonstrate that
  the electron-phonon coupling promotes the trapping of charge
  carriers, the disappearance of the orbital polaron pattern and the
  breakdown of ferromagnetism at the CMR transition. The role of
  different superexchange interactions is explored.
\end{abstract}


\end{frontmatter}

Manganese oxides with perovskite structure, such as $\rm
La_{1-x}[Sr,Ca]_{x}MnO_3$, are examples of highly correlated electron
systems near the metal-insulator transition.  In these materials
strong Coulomb interaction $U$ and Hund's rule coupling $J_h$
introduce a spin background and affect the charge mobility via
double-exchange~\cite{Ze51b}. Delocalisation of the $e_g$ valence
electrons favours ferromagnetism (FM) because the kinetic energy of an
itinerant carrier is minimised when all the core ($t_{2g}$ electron)
spins are aligned. A competing tendency towards localisation comes
about through the electron-phonon (EP) interaction $g$, which
stabilises a local distortion $q_\delta$ of the oxygen octahedron
surrounding each manganese ion. At every Mn site two Jahn-Teller (JT) modes
of $E_g$ symmetry, $q_\theta$ and $q_\varepsilon$, couple to the
orbital degree of freedom of the $e_g$ electrons. In addition, a
breathing-mode $q_{a_1}$ is sensitive to the local density of doped
holes. It is the novelty of the manganites that even at high densities
$x$ carriers can be trapped by self-induced lattice distortions, which
provides an efficient additional mechanism for the suppression of the
resistivity at the CMR transition~\cite{Mi98}.

Based on a recently derived microscopic model for doped manganites, 
\begin{eqnarray}\label{model}
  {H} & = & {H}_{\text{double-exchange}} 
  +{H}_{\text{spin-orbital}}^{\text{2nd order}}
  +{H}_{\text{electron-JT}}\nonumber{}
  \\ & & +{H}_{\text{hole-breathing}}
  +{H}_{\text{phonon}}, 
\end{eqnarray}
which restricts the local electronic Hilbert space to the spin-$2$
orbital doublet $^{5}E$ [Mn$^{3+}$ with $t_2^3(^4A_2)e$] and the
spin-$\frac{3}{2}$ orbital singlet $^{4}A_2$ [Mn$^{4+}$ with $t_2^3$]
but includes the full quantum dynamics of charge, spin, orbital and
lattice degrees of freedom (for details see Ref.~\cite{WF02l}), we
study the properties of a small manganite cluster in the physically
most interesting doping regime $x=0.25$ using exact diagonalisation
techniques. In the present numerical work we focus on the role of
second order superexchange interactions $\propto t^2/U$ and compare
the cases $U=6$~eV and $U\to\infty$, where only antiferromagnetic
(AFM) spin-exchange terms $\propto t^2/J_h$ survive. Taking realistic
values for the hopping $t=0.4$~eV, $t/t_{\pi}=3$, the Hund's coupling
$J_h = 0.7$~eV, and the phonon frequency $\omega=70$~meV, we solve the
model~(\ref{model}) for increasing EP interaction $g$.

\begin{figure}
  \includegraphics[width=\linewidth]{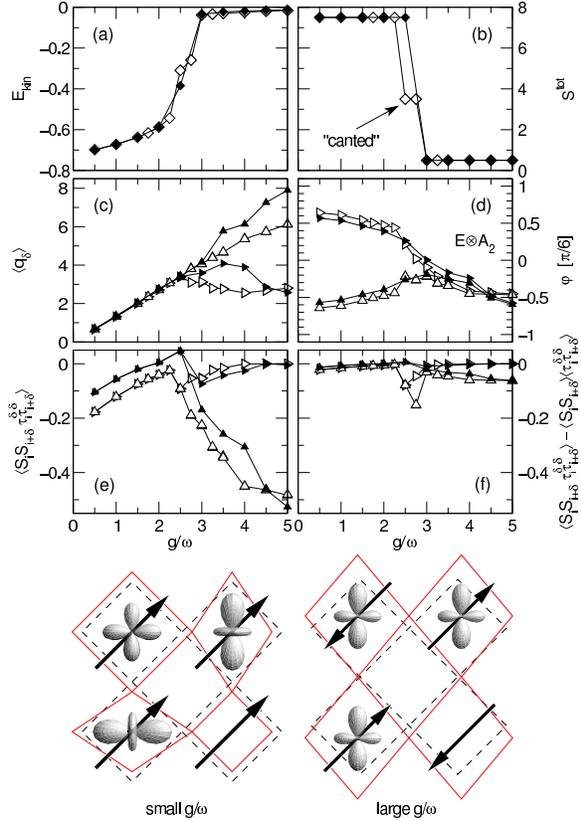}
  \caption{Kinetic energy (a), total spin of  the ground state (b),
    bond-length $\langle q_{x/y}\rangle$ along $x$ and $y$ direction
    (c), orbital orientation $\varphi$ in the neighbourhood of a hole
    (d), expectation value of the spin-orbital interaction (e), and
    spin-orbital correlations (f) are given as a function of the rescaled
    EP coupling $g/\omega$.  Open (filled) symbols refer to $U=6$
    ($U=\infty$) and right (upward) triangles correspond to $\delta=x$
    ($\delta=y$) in the panels (c)-(f). The lowest panel schematically
    visualises the lattice displacements and spin/orbital patterns at
    small and large EP interactions.}\label{fig1}
\end{figure}

The behaviour of selected quantities characterising the ground state
properties in different electron-electron and EP coupling regimes are
shown in Fig.~\ref{fig1}. The main results are the following:

(i) Increasing EP interaction leads to a polaronic reduction of the
kinetic energy (Fig.~\ref{fig1}~(a)) which in turn destroys the FM
state (Fig.~\ref{fig1}~(b)).  For finite $U$ the system goes through a
intermediate canted state, since second order FM and AFM interactions
compete.

(ii) The change from FM to AFM spin correlations is accompanied by a
finite net lattice distortion $\langle q_y - q_x \rangle\ne 0$ in the
$x$-$y$-plane (Fig.~\ref{fig1}~(c)). The associated fluctuation
$\langle q_y^2\rangle-\langle q_y\rangle^2$ shows a kink near the FM
to AFM transition point~\cite{WF02l,BBKLCN98}.

(iii) Orbital correlations change with increasing $g$:
Fig.~\ref{fig1}~(d) displays the behaviour of the $E\otimes A_2$
states $(\cos\varphi|\theta\rangle_i +
\sin\varphi|\varepsilon\rangle_i)\otimes|a_2\rangle_j$ on a bond
$\langle ij \rangle$, where $|\theta\rangle$, $|\epsilon\rangle$ and
$|a_2\rangle$ denote the orbital doublet and singlet states,
respectively. We observe the expected orbital polaron pattern, as
long as the itinerant hole is mobile, and static uniform orbital
order, as soon as the lattice is distorted. The orbital correlations
do not depend much on the value of $U$.

(iv) The expectation value of the spin-orbital interaction is
dominated by the orbital correlations below and above the transition (cf. Fig.~\ref{fig1}~(e)).
Moreover, it is important to notice that the coupling between spin and
orbital degrees of freedom, which is measured by $\langle{\bf
  S}_{i}{\bf S}_{i+\delta}\,
\tau^{\delta}_{i}\tau^{\delta}_{i+\delta}\rangle - \langle{\bf
  S}_{i}{\bf S}_{i+\delta}\rangle
\langle\tau^{\delta}_{i}\tau^{\delta}_{i+\delta}\rangle$ and shown in
Fig.~\ref{fig1}~(f), is almost negligible except for the canted state
near the FM-AFM transition point at finite $U$. This observation
corroborates the decoupling scheme proposed by Khaliullin and
Oudovenko~\cite{KO97} for the use with Greens function approaches.

The schematic drawing in the lowest line of Fig.~\ref{fig1}
summarises, to some extent, the evolution of the lattice, spin and orbital
correlations with increasing EP interaction. It is evident that the
interplay between the double-exchange, superexchange-orbital, and
electron-lattice interactions is the reason for the rich phase diagram
of the manganites.

\end{document}